\pdfoutput=1
\documentclass[iop]{emulateapj}
\usepackage{apjfonts}
\newcommand{\be}{\begin{equation}}
\newcommand{\ee}{\end{equation}}
\newcommand{\bea}{\begin{eqnarray}}
\newcommand{\eea}{\end{eqnarray}}
\newcommand{\Msun}{M_{\odot}}

\def\kms{\ {\rm km\, s}^{-1}}

\def\msun{\,M_{\odot}}
\def\mhalo{\,M_{\rm halo}}

\def\Sst{\Sigma_{\ast}}
\def\rhost{\rho_{\ast}}
\def\Mst{M_{\ast}}
\def\rcool{r_{\rm cool}}
\def\Zsol{Z_{\odot}}

\shortauthors{CONROY, VAN DOKKUM, \& KRAVTSOV}
\shorttitle{PREVENTING STAR FORMATION WITH LATE-TIME STELLAR HEATING}

\begin{document}

%---------------------------------------------------------%
\title{Preventing Star Formation in Early-Type Galaxies with Late-Time
  Stellar Heating}
%---------------------------------------------------------%

\author{Charlie Conroy\altaffilmark{1,2},
  Pieter G. van Dokkum\altaffilmark{3},
  Andrey Kravtsov\altaffilmark{4,5} }

\altaffiltext{1}{Department of Astronomy \& Astrophysics, University
  of California, Santa Cruz, CA, USA}
\altaffiltext{2}{Department of Astronomy, Harvard University,
  Cambridge, MA, USA}
\altaffiltext{3}{Department of Astrophysical
  Sciences, Yale University, New Haven, CT, USA}
\altaffiltext{4}{Department of Astronomy \& Astrophysics, 5640 South
  Ellis Ave., The University of Chicago, Chicago, IL 60637} 
\altaffiltext{5}{Kavli Institute for Cosmological Physics and 
  Enrico Fermi Institute, 5640 South Ellis Ave., The University of
  Chicago, Chicago, IL 60637}

\slugcomment{Submitted to ApJ}

\begin{abstract}

  We revisit previous suggestions that the heating provided by the
  winds of dying low-mass stars plays an important role in preventing
  star formation in quiescent galaxies.  At the end of their
  asymptotic giant branch phase, intermediate and low-mass stars eject
  their envelopes rapidly in a super-wind phase, usually giving rise
  to planetary nebulae.  In spheroidal galaxies with high stellar
  velocity dispersions, the interaction of these ejected envelopes
  with the ambient diffuse gas can lead to significant, isotropic and
  steady-state heating that scales as $\dot{M}_\ast\sigma_\ast^2$.  We
  show that cooling of the central regions of the hot diffuse halo gas
  can be delayed for a Hubble time for halos more massive than
  $\sim10^{12.5}\msun$ at $0<z<2$, although stellar heating alone is
  unlikely to forestall cooling in the most massive clusters at
  $z=0$. This mechanism provides a natural explanation for the strong
  trend of galaxy quiescence with stellar surface density and velocity
  dispersion.  In addition, since the ejected material will thermalize
  to $kT\sim\sigma_\ast^2$, this mechanism provides an explanation for
  the observed similarity between the central temperature of the hot
  diffuse gas and $\sigma_\ast^2$, a result which is not trivial in
  light of the short inferred cooling times of the hot gas.  The main
  uncertainty in this analysis is the ultimate fate of the stellar
  ejecta. Preventing accumulation of the ejecta in the central regions
  may require energy input from another source, such as Type Ia
  supernovae.  Detailed simulations of the interaction of the stellar
  wind with the ambient gas are required to better quantify the net
  effect of AGB heating.

\end{abstract}


%---------------------------------------------------------%

\section{Introduction}
\label{s:intro}

Galaxies with little-to-no star formation are common today and have
been discovered at remarkably early epochs \citep[$z=2-4$;
e.g.,][]{Daddi05, Kriek06a, Onodera12, Whitaker13}.  The existence of
these quiescent galaxies poses two major questions to galaxy formation
models: what causes the cessation of star formation (quenching) and
what {\it maintains} the low observed star formation rates over
cosmological timescales.  The second issue is challenging because
massive quiescent galaxies are embedded in dark matter halos
containing hot diffuse gas that, in the absence of a heat source to
offset radiative cooling, will eventually form new stars
\citep[e.g.,][]{Fabian84, Croton06}.  Hydrodynamic simulations with
standard recipes for gas cooling and energy feedback from Type II
supernovae are generally unable to produce realistic quiescent
galaxies \citep[e.g.,][]{Martizzi14}.

A variety of heating sources have been proposed in order to offset the
radiative losses of the diffuse gas, although by far the most popular
is some form of energetic feedback from an active galactic nucleus
(AGN).  Evidence that supermassive black holes do {\it something} to
the diffuse gas in low redshift clusters is ubiquitous, including the
existence of giant radio cavities, sound waves, non-thermal pressure
support, etc. \citep[see][for a recent review]{McNamara07}.  However,
it has been difficult to convincingly demonstrate that this activity
is directly responsible for heating the diffuse gas.

Recently it has become clear that galaxy quenching is more strongly
correlated with their central stellar densities and velocity
dispersions, than their overall stellar mass
\citep[e.g.,][]{Kauffmann03b, Franx08, Bell12, Cheung12, Fang13}. This
is particularly striking at $z\sim2$, where $\sim50$\% of galaxies
dwith $M_\ast>10^{11}\Msun$ are quiescent and $\sim50$\% are star
forming \citep{Brammer11} and the stellar density is an excellent
predictor of star formation rate \citep{Franx08}.  This suggests that
the stars may play an important role in either quenching galaxies or
maintaining the low observed star formation rates.  Morphological
quenching is one possible mechanism, in which the high velocity
dispersion stabilizes a gaseous disk against fragmentation
\citep[e.g.,][]{Martig09}.  However, in this scenario the cooling of the
hot gas would likely result in the accumulation of giant gas disks
that would eventually become unstable.

In this paper we consider the possibility that the old stars {\em
  themselves} are responsible for heating the gas, thus potentially
obviating the need for a separate mechanism (such as AGN feedback) to
explain the strong correlation between high stellar densities and the
absence of star formation. We will focus on mass loss by AGB stars,
both in isolation and in tandem with the effects of Type Ia supernova
explosions.  In this scenario the heating rate scales directly with
both the local stellar density and the stellar velocity dispersion
squared (the latter arises because of the thermalization of the
stellar winds).  These heating sources are an inevitable consequence
of stellar evolution; the main questions are whether there is enough
energy to offset radiative cooling and the extent to which wind
material accumulates in the central regions of galaxies.

It has long been recognized that dying low-mass stars and Type Ia SNe
may contribute to the heating of diffuse gas in quiescent galaxies
\citep[e.g.,][]{Mathews71, Lake84, Mathews90, Ciotti91, Matsushita01}.
Much of the early work focused on questions relating to the origin of
hot gas and cooling flows in massive systems. Here we re-examine these
heating mechanisms in the context of current galaxy formation models
to determine whether they are sufficiently powerful to explain the low
observed star formation rates in quiescent galaxies, and especially in
the population of compact quiescent galaxies at high redshift.

%---------------------------------------------------------%

\section{Basic Arguments}
\label{s:methods}

\begin{figure}[!t]
\center
\includegraphics[width=0.52\textwidth]{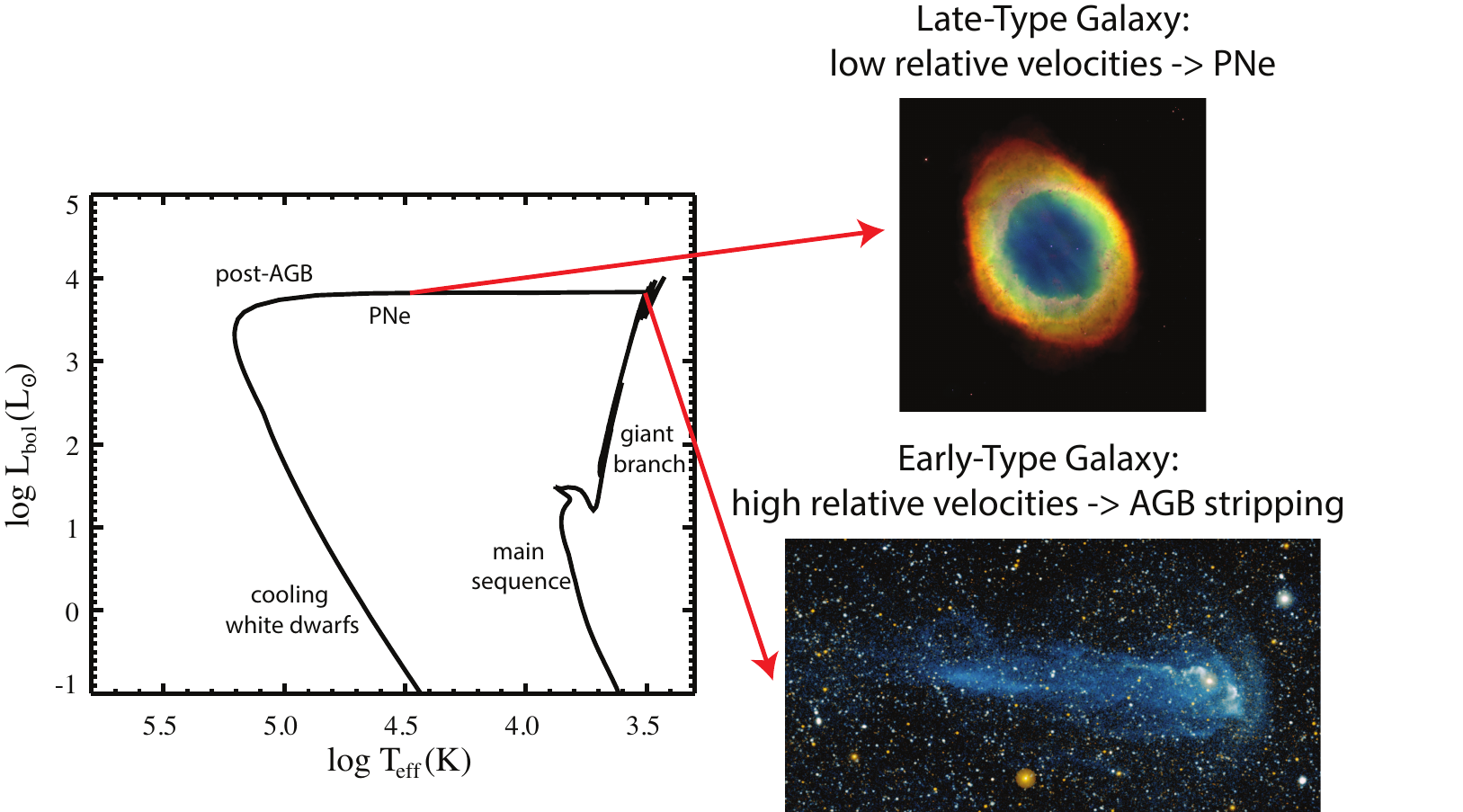}
\vspace{0.1cm}
\caption{Schematic illustration of the thermalization of winds from
  low mass stars.  When AGB stars move with low velocity relative to
  the surrounding gas, ejected material remains near to the central
  star, creating planetary nebulae (such as the Ring Nebula shown in
  the top right panel from the Hubble Heritage Team).  However, when
  AGB stars move with high velocities relative to the ambient gas, the
  ejected matter is swept away by ram pressure such as seen in the UV
  image of Mira in the bottom right panel; image credit:
  NASA/JPL-Caltech/C. Martin/M. Seibert.  Though Mira is in the Milky
  Way, its velocity of $130\,\kms$ relative to the ambient gas
  \citep{Martin07} is a configuration expected to be common in the
  central regions of early-type galaxies.  In this case the energy
  required to decelerate the ejected material is transfered to the
  background gas, which can result in a net heating of the gas with
  $\sigma_\ast^2\gtrsim T$.}
\label{fig:cartoon}
\end{figure}

\subsection{Stellar Wind Heating}
\label{s:basic}

We begin by considering the heating due to late-time stellar mass loss
from evolved low mass stars.  Consider a region with stellar density,
$\rhost$. A stellar population of initial mass $\Mst$ and age $t$
loses mass at a rate that can be approximated as
$\dot{M}_{\ast}\approx8\, (\Mst/\Msun)\,(t/{\rm yr})^{-1.25}$ $\Msun$
yr$^{-1}$ for ages $>10^7$ yrs, and a \citet{Chabrier03} IMF
\citep[estimated from the models of][]{Conroy09a}.  Much, perhaps
most, of the mass lost by old stellar populations is due to a single
phase of stellar evolution: near the end of the AGB phase, stars
undergo a catastrophic mass-loss event in which they shed their
remaining envelope \citep{Iben83}.\footnote{Whether or not the
  majority of the mass-loss occurs very rapidly or slowly (as might be
  the case for RGB mass-loss) is immaterial for our purposes because
  in both cases the background gas must do work to decelerate the
  ejected material to zero relative velocity.} The ejection of the
envelope often leads to the formation of a planetary nebula.  The cold
gaseous envelope shed in this stage shares the velocity of the parent
star relative to the ambient gas and will thus interact with the gas
\citep[e.g.,][]{Mathews90}.  The work done by the background gas to
decelerate the circumstellar shell may result in a net heating of the
background gas.  See Figure \ref{fig:cartoon} for a schematic
illustration.

In addition to heating the surrounding ambient gas, some energy must
also go into heating the stellar ejecta in order for late-time stellar
heating to be effective.  In fact, when $T>\sigma_\ast^2$, the
interaction of the AGB ejecta with the ambient gas will result in net
cooling, not heating.  However, if cooling brings the temperature of
the ambient gas substantially below $\sigma_\ast^2$, the
thermalization of AGB ejecta will heat the gas back to
$\sigma_\ast^2\sim T$.  In the calculations that follow we assume the
maximum heating rate for the case when $T\ll \sigma_\ast^2$.  In
\S\ref{s:disc} we mention the possibility that AGB heating may
ultimately be responsible for the observed similarity between $T$ and
$\sigma_\ast^2$ for groups and clusters.

If the entire kinetic energy of the lost gas were transferred to
internal energy of the ambient gas, the heating rate per unit volume
would be
\begin{equation}
\dot{e}_{\rm heat}= \frac{1}{2}\frac{\dot{M}_{\ast}\sigma_{\ast}^2}{V}= \frac{1}{2}\frac{8\Mst \sigma_{\ast}^2 }{Vt^{1.25}}=\frac{4\rhost \sigma_{\ast}^2}{t^{1.25}}.
\end{equation}
Here $V$ is a volume element and $\sigma_{\ast}$ is the three-dimensional
velocity dispersion of stars.

The cooling rate per unit volume of the ambient gas of number density
$n_g$ is $\dot{e}_{\rm cool}=n_g^2\Lambda(T,Z)$, where $\Lambda(T,Z)$
is the cooling function depending on temperature and metallicity.
Herein we adopt the cooling function of \citet{Sutherland93}.

The ratio of heating to cooling rate is then:
\begin{eqnarray}\label{eqn:ratio}
\frac{\dot{e}_{\rm heat}}{\dot{e}_{\rm cool}}&=&
\frac{{4\rhost} \sigma_{\ast}^2}{t^{1.25}n^2\Lambda(T,Z)}\approx \, 10
{\rho_{\ast}}_1\sigma_{\ast,300}^2n^{-2}_{g,-2}\Lambda^{-1}_{-23} t^{-1.25}_{9.7},
\end{eqnarray}
where $\rho_{\ast,1}$ is in units of $\Msun\,{\rm pc^{-3}}$,
$\sigma_{\ast,300}$ is the velocity dispersion in units of $300$ km/s,
$n_{g,-2}$ is the ambient gas number density in units of $10^{-2}\,\rm
cm^{-3}$, $\Lambda_{-23}$ is the cooling function in units of
$10^{-23}\rm\, erg\,s^{-1}\,cm^{3}$ and $t_{9.7}$ is the stellar
population age in units of $5\times10^9$ yrs, which is typical for an
average age of stellar populations of both the inner regions of
$z\sim0$ disk galaxies and ellipticals.  For the temperature range
relevant for ambient halo gas, $T\sim 10^5-10^7$ K of metallicity,
$Z\sim 0.01-1.0Z_{\odot}$, $\Lambda$ ranges from $\sim 10^{-23}$ to
$\sim 10^{-22}$ \citep[e.g.,][]{Sutherland93}.

Finally, the AGB heating expressed as a total power available is
\noindent
\be
\dot{E}_{\rm AGB} \approx 5\times 10^{40} \, M_{\ast,11}\,
\sigma_{\ast,300}^2\, t_{9.7}^{-1.25}\,\, {\rm erg\,s}^{-1},
\ee
\noindent
where $M_{\ast,11}$ is the stellar mass in units of $10^{11}\Msun$.

We infer that the heating rate from stellar mass loss could, in this
optimistic scenario, exceed the cooling rate by a factor of $\sim10$
for massive compact galaxies.  

In these calculations we have ignored the fact that stellar ejecta
will accumulate in the central regions.  If left unchecked, the
growing central gas density will increase the cooling rate of the
ambient gas, eventually to the point where cooling may overwhelm AGB
heating.  This issue is addressed in the next section.

\subsection{Type Ia SNe Heating}
\label{s:sne}

We can compare the energy available in wind thermalization to the
total energy available from another significant late-time stellar
heating source: Type Ia SNe explosions.  The delay time distribution
of type Ia SNe has been estimated by \citet{Maoz12} to be
$\approx3\times10^{-2}\,t^{-1.1}$ SNe per year per $M_{\ast,11}$ where
$t$ is in Gyr.  If we adopt an energy deposition per event of
$10^{51}$ erg, then the power available from Ia SNe is:
\noindent
\be 
\dot{E}_{\rm Ia} \approx
14\times10^{40}\, M_{\ast,11}\,t_{9.7}^{-1.1}\,\, {\rm erg\,s}^{-1}.
\ee
\noindent
Notice that the time and mass dependence of the two forms of heating
are very similar and for massive galaxies they are comparable.  Both
sources will generally work in tandem to heat the gas, with their
relative importance a function of velocity dispersion: the ratio of
these energy sources scales as $\sigma_\ast^2$, such that in galaxies
with velocity dispersions less than $\sim300$ km s$^{-1}$ the heating
due to Ia SNe is much larger than the heating due to wind
thermalization.  We expect AGB heating to dominate in massive and
compact galaxies.

Furthermore, notice that the binding energy of the accumulated
wind material evolves as:
\be
\dot{E}_g = \frac{G\,M_\ast\,\dot{M}_\ast}{R_\ast} 
  \approx 9\times10^{40} \, M_{\ast,11} \,  t_{9.7}^{-1.25}\, R_{\ast,1}^{-1}\, {\rm erg\,s}^{-1},
\ee
where $R_{\ast,1}$ is the galaxy half-mass size in units of 1 kpc.  The thermal
energy provided by type Ia SNe explosions should therefore be 
sufficient to unbind the accumulated material.

In the rest of this article we will refer to `AGB heating' and
`late-time stellar heating' interchangeably, but a more precise phrase
would be `SNe Ia-assisted AGB heating' to emphasize that the clearing
of accumulated material by Ia SNe is an important aspect of this
scenario.

\begin{figure}[!t]
\center
\includegraphics[width=0.49\textwidth]{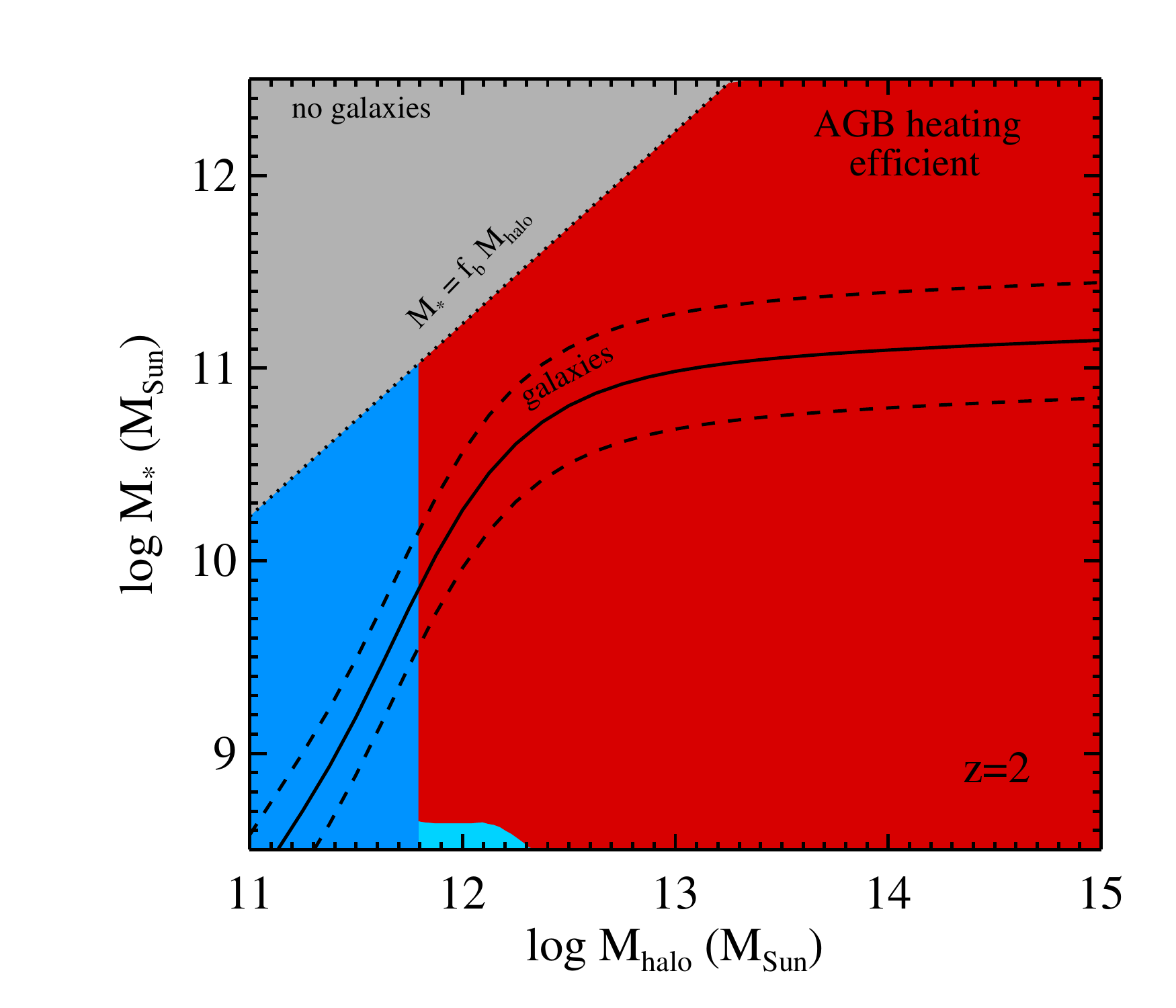}
\includegraphics[width=0.49\textwidth]{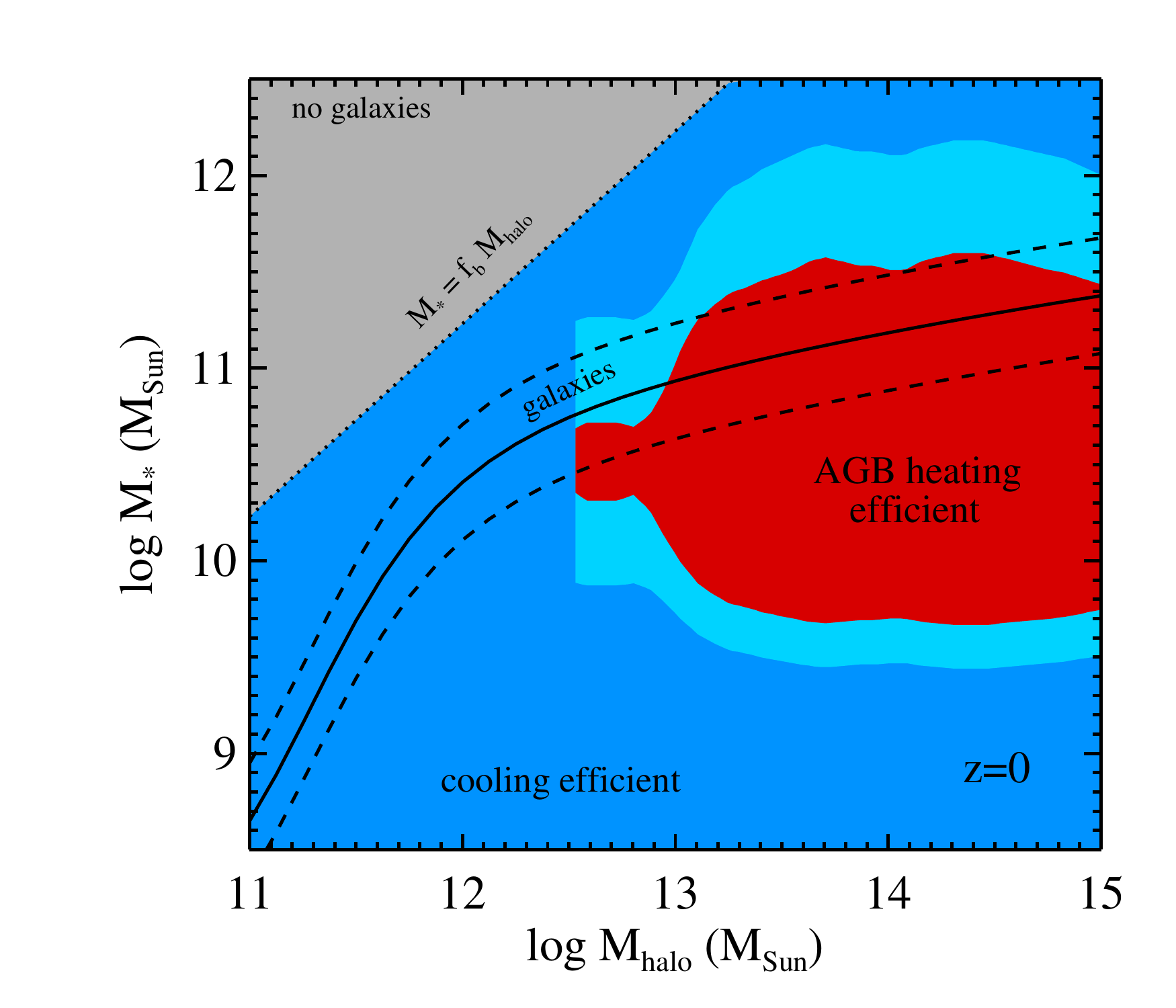}
\caption{Potential efficiency of AGB heating.  This diagram compares
  radiative cooling of diffuse halo gas to AGB heating as a function
  of galaxy stellar mass, $\Mst$, and halo mass, $\mhalo$, at $z=2$
  ({\it top panel}) and $z=0$ ({\it bottom panel}).  Red and blue
  regions denote where AGB heating and radiative cooling dominate,
  respectively.  The light blue region marks where heating and cooling
  balance to within a factor of two.  Galaxies occupy the region
  within the dashed lines \citep{Behroozi13}.  The sharp vertical
  boundary between heating and cooling is a consequence of our
  assumption that halos with $T_{\rm vir}<10^6$ K are always able to
  cool efficiently.}
\label{fig:hvc}
\end{figure}

\begin{figure}[!t]
\center
\includegraphics[width=0.49\textwidth]{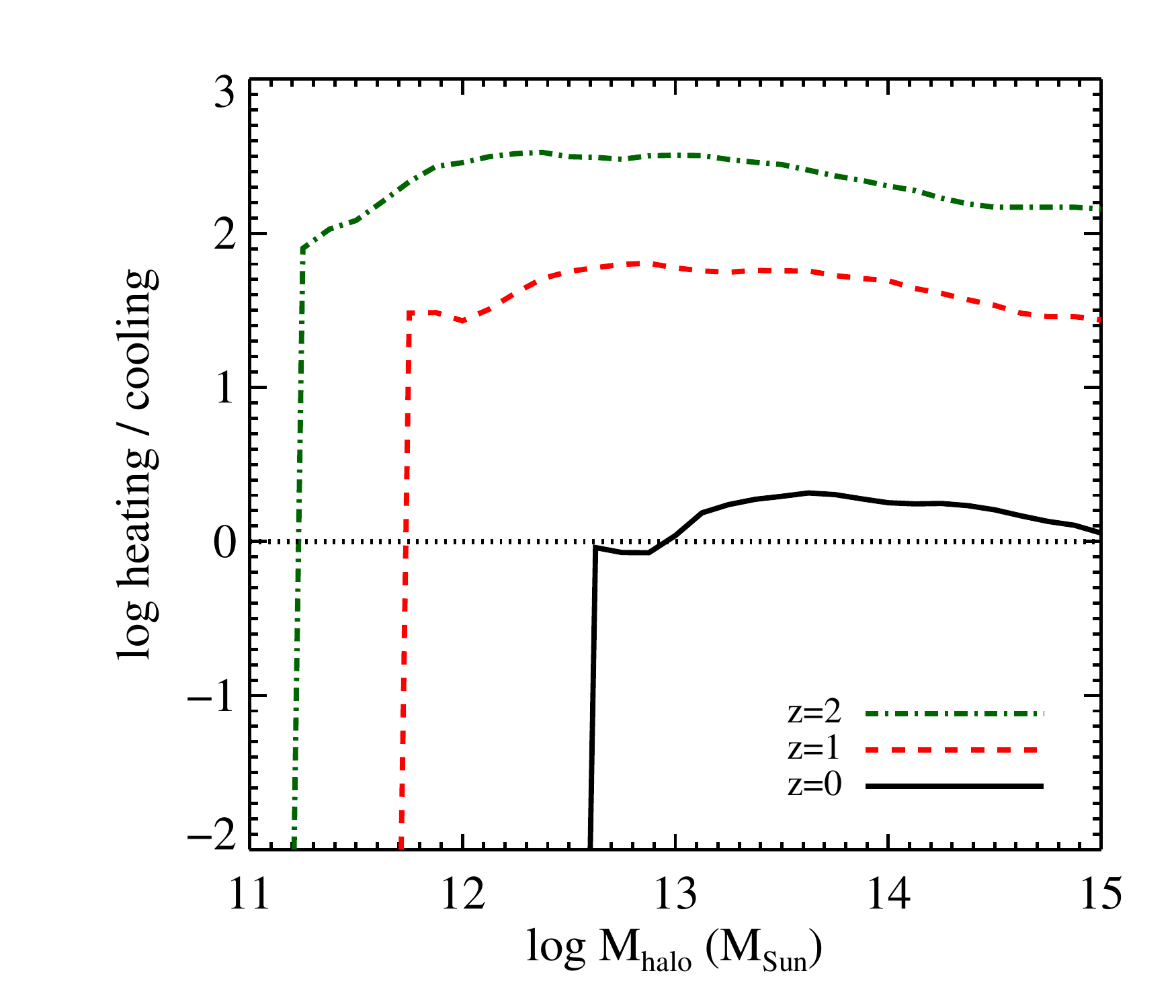}
\includegraphics[width=0.49\textwidth]{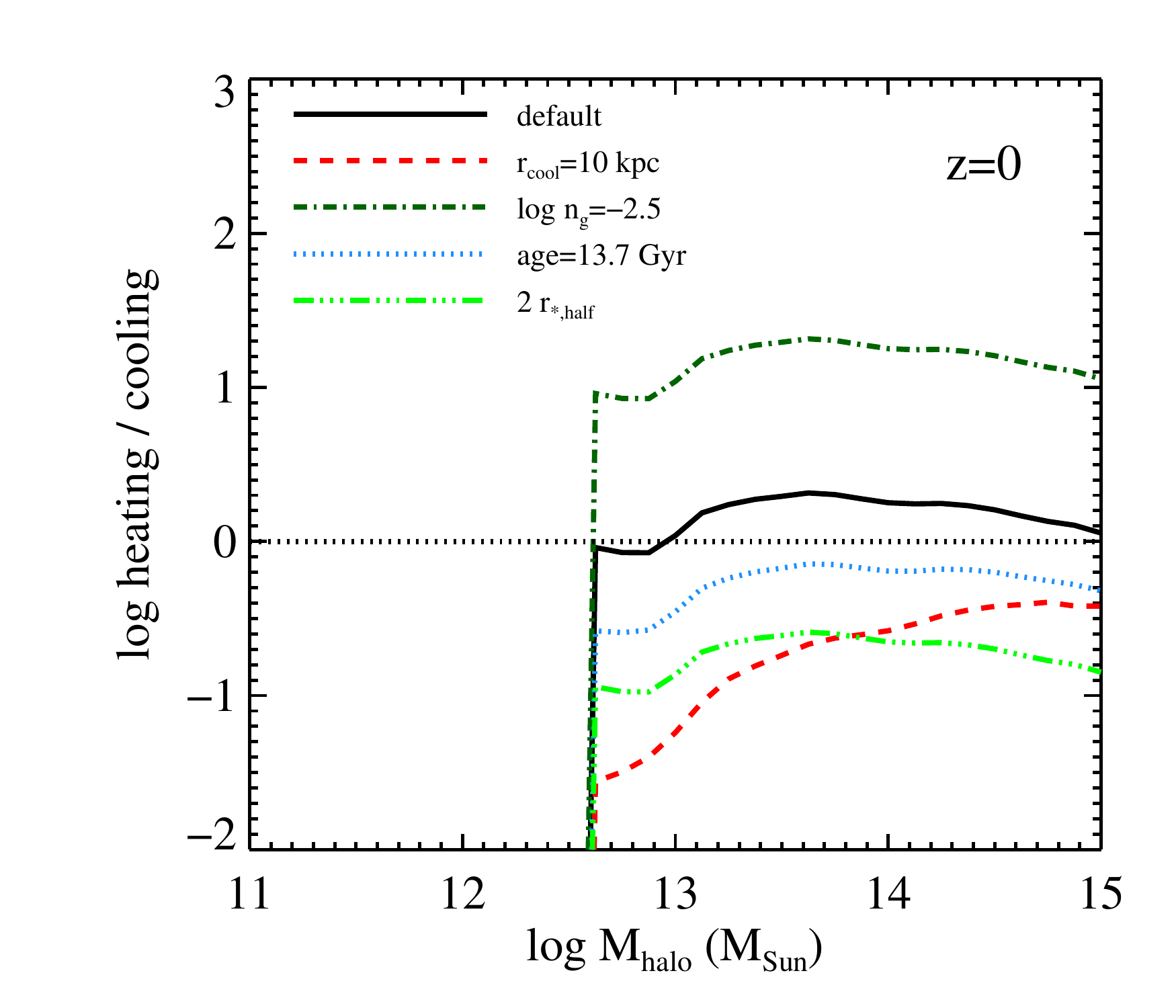}
\caption{Balance between AGB heating and radiative cooling as a
  function of halo mass. At each halo mass the heating and cooling
  rates are computed for the galaxy expected to occupy the halo.  {\it
    Top panel:} Results for our default model assumptions at
  $z=0,1,2$.  {\it Bottom panel:} Results after varying several model
  assumptions at $z=0$.  The variations include fixing the cooling
  radius to 10 kpc and setting $r_{\rm cool}=2\,r_{\ast,\rm half}$,
  decreasing the gas density to $10^{-2.5}$ cm$^{-3}$ and fixing the
  stellar ages to 13.7 Gyr for all galaxies.}
\label{fig:sum}
\end{figure}

\section{Mass and Redshift Dependence of Late-Time Stellar Heating}

\subsection{Expectations}

We now proceed to compute the ratio between AGB heating and radiative
cooling (i.e., Equation \ref{eqn:ratio}) as a function of stellar
mass, $\Mst$, dark matter halo mass\footnote{Halo masses are defined
  according to the \citet{Bryan98} virial definition.  Where necessary
  a normalized Hubble constant of $h=0.7$ is adopted.}, $\mhalo$, and
redshift.  Based on the arguments in the previous section, we assume
that the accumulated AGB ejecta is cleared from the central regions of
the galaxy by type Ia SNe.  Here we are interested in global
quantities, and so we replace $\rhost$ with $\Mst$ and $n_g$ with
$N_g$ (total number of gas particles).  The former is a relatively
well-defined quantity but the latter requires a choice of radius.
Here we adopt a radius of $\rcool=r_{\ast,\rm half}$, where
$r_{\ast,\rm half}$ is the half-mass radius of the stars.  We regard
AGB heating to be effective if it can offset radiative losses within
the cooling radius, $\rcool$.  The cooling radius is the most
important adopted parameter in what follows because the total cooling
rate scales as $\rcool^3$.  We consider alternative definitions of
$\rcool$ below.

For the properties of the galaxies and the stars within them, we make
the following assumptions.  We adopt the $\Mst-\sigma_\ast$ relation
at $z=0$ for early-type galaxies from \citet{Dutton11b}, which is a
complex function but basically scales as
$M_\ast\propto\sigma_\ast^{3.5}$.  We assume that at fixed
$\sigma_\ast$, the stellar mass of galaxies decrease by a factor of
two from $z=0$ to $z=2$ \citep[e.g.,][]{Belli14}.  Note that
observations report a one dimensional velocity dispersion; we multiply
the reported values by $\sqrt{3}$ to approximate the three dimensional
dispersion.  We also adopt the $z=0$ mass-size relation for early-type
galaxies from \citet{Dutton11b} and scale it by $(1+z)^{-1}$ for
higher redshifts \citep{vanderWel14}.  We also must specify a typical
age of the stars in order to set the mass-loss rate.  Here we use the
$\Mst$-stellar age relation at $z=0$ from \citet{Choi14}, and we scale
the ages by the Hubble time as a function of redshift.  We adopt a
minimum age of 2 Gyr as we do not expect quiescent galaxies to have
ages considerably younger than this, by definition.

For the diffuse halo gas we adopt the following model.  We assume the
diffuse gas is isothermal and at the halo virial temperature, $T_{\rm
  vir}$, which for groups and clusters at $z=0$ is accurate to within
a factor of $\sim1.5-2.5$ \citep{Vikhlinin06a, Sun09}.  For halos with
$T_{\rm vir}<10^6$ K we assume that there is no stable hot halo and
that available gas is able to settle into a cold disk. The diffuse gas
has a metallicity of $\frac{2}{3}\,\Zsol$, which is appropriate for
the central regions of clusters \citep{Rasmussen07, Johnson11,
  Sasaki14}.  The choice of the diffuse gas density is a second
(following $\rcool$) key model parameter.  Here we assume that the
ambient gas density is constant within $\rcool$ and with redshift and
equal to $10^{-2}$ cm$^{-3}$.  Empirically this is a good
approximation for groups and clusters at $z=0$ \citep{Vikhlinin06a,
  Sun09}.  For lower mass systems the gas density is not
well-constrained but a variety of arguments, including results from
simulations, favor values of $n\sim 10^{-2}\rm\, cm^{-3}$ or lower
\citep[e.g.,][]{Guedes11,Feldmann13, Agertz13}.  A constant gas
density within $\rcool$ is a reasonable assumption here since $\rcool$
is proportional to $r_{\ast,\rm half}$, and the latter is proportional
to the halo virial radius \citep{Kravtsov13}.  The cooling radius in
our model is therefore tracking a fixed fraction of the halo virial
radius, and assuming a self-similar gas density profile, one might
expect the mean density within $\rcool$ to be constant with mass and
redshift.

With the assumptions above we are now ready to compute the balance
between AGB heating and radiative cooling as a function of $\Mst$ and
$\mhalo$ at various redshifts.  We show this balance in Figure
\ref{fig:hvc}.  Red regions in this diagram indicate where AGB heating
is effective at offsetting radiative losses.  We also show the
expected locations of galaxies based on the abundance matching
technique \citep{Behroozi13}.

The main result of this paper is that AGB heating can be an efficient
mechanism to offset radiative losses from diffuse gas over much of the
$\Mst-\mhalo$ plane, and certainly in the regime where quiescent
galaxies reside (roughly above the bend in the galaxy $\Mst-\mhalo$
relation).  This mechanism is more efficient at high redshift for
several reasons: 1) galaxies are more compact at higher redshift, and
so the adopted cooling radius is smaller; 2) at fixed stellar mass
galaxies have higher velocity dispersions at higher redshift; and 3)
the stellar mass-loss rates are higher at high redshift.  We emphasize
here that we are considering the ability of AGB heating to offset
cooling of {\it hot diffuse gas}.  At low halo masses the majority of
the gas is expected to be cool and dense, and so we do not expect AGB
heating to compete with radiative cooling (see also \S\ref{s:obs}
below).

AGB heating is unlikely to be effective at shutting down star
formation in the cold interstellar medium (ISM).  The densities of the
cold ISM are several orders of magnitude higher than the limit where
post-AGB heating is effective at offsetting radiative cooling (see
Figure \ref{fig:ngas} below).  Furthermore, in dynamically cold
systems the relative velocities between the gas and the stars are in
general low, which means that the amount of energy transfered from the
circumstellar envelope to the gas will be low.

Figure \ref{fig:sum} shows the balance between AGB heating and
radiative cooling as a function of $\mhalo$.  At each halo mass the
heating and cooling rates are computed for the galaxy expected to
occupy the halo.  The top panel shows our default model assumptions at
$z=0,1,2$ while the bottom panel shows results after varying several
model assumptions.  At $z=1,2$ the heating rates are very high both
because the stellar ages are young and because the stellar densities
and velocity dispersions are high at a fixed $\Mst$.  As expected,
$\rcool$ and $n_g$ have a significant effect on our results which is
not surprising because the total cooling rate scales as
$\rcool^3\,n_g^2$.

\subsection{Comparison to Observations}
\label{s:obs}

\begin{figure}[!t]
\center
\includegraphics[width=0.49\textwidth]{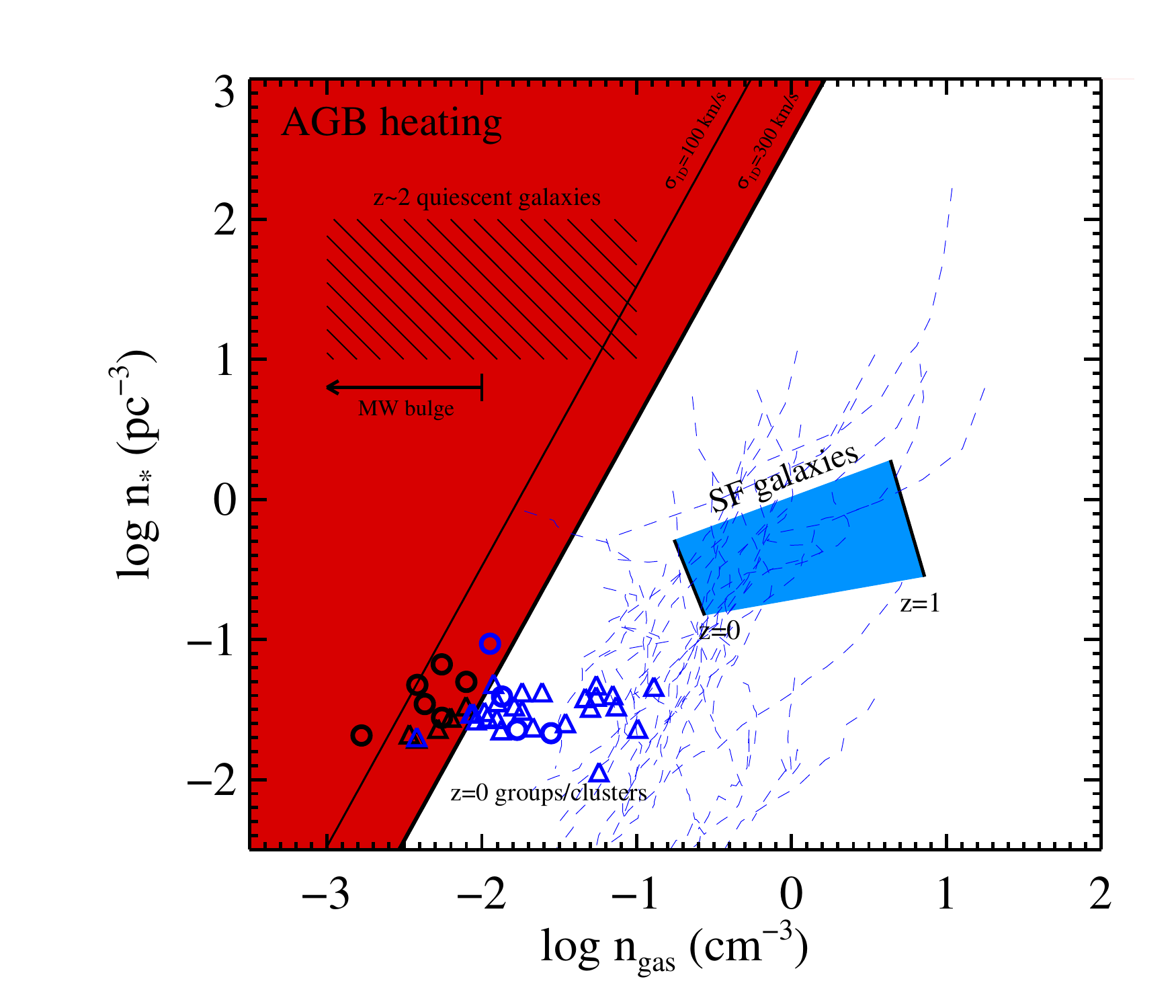}
\caption{Local efficiency of AGB heating compared to radiative cooling
  as a function of gas and stellar density (red denotes regions where
  heating is dominant).  Also shown is the expected location of star
  forming (SF) galaxies at $0<z<1$ and the approximate locus of
  quiescent galaxies at $z=2$.  For the latter, the range of gas
  densities shown is the expected central gas densities for gas in
  hydrostatic equilibrium with dark matter halos where the gas is at
  the halo virial temperature.  Dashed lines are radial gas and
  stellar profiles of disk galaxies from the THINGS survey
  \citep{Leroy08}.  Also shown is the location of the central $\sim10$
  kpc of local galaxy groups \citep[][open circles]{Sun09} where
  H$\alpha$ imaging \citep{McDonald11} shows evidence for either no
  emission (black symbols) or extended emission (blue symbols), and
  galaxy clusters from \citet[][triangles]{Haarsma10} color-coded by
  the central cooling time ($>10^{9.5}$ yr in black and younger in
  blue).  }
\label{fig:ngas}
\end{figure}

Figure \ref{fig:ngas} shows the region where AGB heating is effective
at a local level, in the gas-stellar density plane.  In this diagram
we show the expected locus of $z\sim2$ quiescent galaxies, the
observed locations of central galaxies within $z=0$ groups and
clusters, the Milky Way bulge, stellar and gas density profiles from
the THINGS survey, and expected locations of typical star-forming
galaxy disks at $0<z<1$.  We estimate the latter by adopting mass-size
relations for disks from \citet{vanderWel14}, the mass-SFR relation
from \citet{Whitaker12}, and a fixed gas depletion timescale of 2 Gyr
(defined as $\tau_{\rm dep}\equiv\Sigma_{\rm gas} / \Sigma_{\rm
  SFR}$).  We also assume that the gas disk scale length is twice that
of the stars and the scale heights of the gas and disk are 10\% of the
corresponding scale lengths.

The number density of the hot halo gas in the inner few kpc of Milky
Way-like galaxies is expected to be close to $n\sim 10^{-2}\rm\,
cm^{-3}$ \citep[e.g.,][]{Guedes11,Feldmann13}.  Figure \ref{fig:ngas}
shows that for these fiducial values AGB heating can offset cooling
for stellar densities $\rhost\gtrsim 1\,\Msun\,\rm pc^{-3}$ assuming
$\sigma_\ast=100\kms$. For disk galaxies of a typical scale height
$h_{\ast}\approx 200$ pc \citep{Bovy13}, such densities would
correspond to a surface density of $\Sst\sim 4\rhost h_{\ast}\sim
800\,\Msun\,\rm pc^{-2}$, which is close to the surface densities that
separate blue star forming galaxies at $z\approx 0$ from the galaxies
in the green valley and red sequence \citep{Fang13}.

Another way to explore the balance between stellar ejecta heating and
radiative cooling would be to compare $\dot{M}_{\ast}\sigma_{\ast}^2$
directly to the X-ray luminosity within a radius where the cooling
times are short.  \citet{Matsushita01} and \citet{Nagino09} have made
precisely this comparison for $\sim50$ early-type galaxies in field
and cluster environments.  These authors find that in nearly all cases
(excluding cD galaxies in massive clusters) the kinetic heating rate
by stellar mass loss is comparable to or greater than the observed
X-ray luminosity.  Furthermore, \citet{Nagino09} find the intriguing
result that the temperature gradient in the ambient gas is correlated
with the ratio of $L_B\sigma^2$ to X-ray luminosity.  This suggests
that stellar ejecta and SNe Ia heating, both of which should correlate
with $L_B\sigma^2$ (though in different ways), play an important role
in shaping the properties of the ambient gas.

Massive quiescent galaxies at $z\sim2$ are remarkably compact, with
typical sizes of $\sim1$ kpc and one dimensional velocity dispersions
of $\gtrsim300\kms$ \citep[e.g.,][]{vanDokkum09, Belli14}.  The
implied stellar densities exceed 10 $\Msun\,{\rm pc^{-3}}$. For these
galaxies, the heating rate of the diffuse gas due to AGB wind
deceleration may exceed radiative losses by factors of $10-100$.  In
fact, at the extreme end of stellar densities and velocity dispersions
observed in high redshift quiescent galaxies, AGB heating may disrupt
even the cold ISM and might therefore play a role in shutting down
star formation in these extraordinary systems.

The mechanism proposed here may explain the results reported in
\citet{Salim12}, who found that low-level star formation on the
optically-defined red sequence was confined mostly to S0 galaxies -
true ellipticals rarely showed UV signatures of star formation.  AGB
heating in S0s will not be isotropic and the mass-loss will be largely
confined to a plane, which may actually fuel the observed low-level
star formation.

%---------------------------------------------------------%

%\vspace{2cm}

\section{Discussion \& Conclusions}
\label{s:disc}

The mechanism described here is a necessary consequence of stellar
evolution, yields a direct, steady-state, and approximately isotropic
heating of the gas, and provides a natural explanation for the
observed tight correlations between the stellar density of galaxies
and their star formation rate \citep[e.g.,][]{Kauffmann03b, Franx08,
  Bell12, Cheung12, Fang13}.  In all these respects it is a viable
alternative to AGN feedback, which is the leading proposed mechanism
for maintaining the low star formation rates in quiescent galaxies
\citep{Croton06}. We emphasize, however, that AGB heating likely
cannot explain why star formation ceases in the first place (except
perhaps in the very densest and highest velocity dispersion galaxies):
it may be that the star forming progenitors of quiescent galaxies
simply consume all of the available cold gas \citep{Feldmann15}, or
that AGN feedback is a source of initial heating and/or removal of
cold gas.  Late-time stellar heating may then provide the long-term
``maintenance'' to keep dead galaxies dead.

There are other potential heating processes associated with old
stellar populations.  The most widely discussed such process is
heating by type Ia supernovae \citep[e.g.,][]{Mathews90, Ciotti91}. As
shown in \S\ref{s:basic}, for massive galaxies the heating rate from
SNe Ia is comparable to the AGB heating rate.  The scaling with
stellar population age and mass is similar for both heating
mechanisms.  However, Ia heating is insensitive to the velocity
dispersion whereas AGB heating scales as $\sigma_\ast^2$.  SNe Ia
heating will therefore dominate AGB heating at low velocity
dispersions, while AGB heating will dominate in massive compact (high
dispersion) galaxies.  In this article we have employed Ia explosions
to clear the accumulated AGB ejecta.  In reality they of will also
likely directly heat the ambient hot gas.  It is beyond the scope of
the present work to consider this second energy source in detail
\citep[see e.g.,][and references therein for discussion]{Ciotti91,
  David06, Negri14}.  Of course, any self-consistent model of galaxy
evolution should include all heating sources, including type Ia SNe
and AGB heating.

AGB heating may also provide a natural interpretation of the observed
central temperatures of the hot gas in groups and clusters.  X-ray
observations of these systems has revealed that the temperature
profiles drop in the central regions by a relatively modest amount,
typically by a factor of $\sim1.5-2.5$ relative to the maximum
temperature found at larger radius \citep[e.g.,][]{Vikhlinin06a,
  Sun09}.  Moreover, the central temperature is close to
$\sigma_\ast^2$ \citep{Matsushita01, Nagino09} .  These observations
are surprising in light of the generally short cooling times
($\lesssim 1$ Gyr) found in the central regions.  The lack of vigorous
star formation indicates that this gas is not cooling
catastrophically, but it is not obvious why a generic heating
mechanism, such as AGN or SNe feedback, would maintain a temperature
so close to $\sigma_\ast^2$.  Indeed, many forms of distributed
heating have difficulty reproducing observed temperature profiles
without fine tuning \citep[e.g.,][]{Conroy08b}.  However, heating due
to wind thermalization would naturally lead to near equality between
the specific internal energy of the ambient gas ($\frac{3}{2} kT/\mu
m_p$) and the specific kinetic energy of the stars
($\frac{1}{2}\sigma_\ast^2$).  This can be tested by carefully
comparing the kinematic profile of central galaxies to the temperature
profiles of the hot gas as a function of halo mass.

The potentially significant role of heating due to stellar ejecta
means that it is important to implement and test this mechanism in
simulations of galaxy formation. Inclusion of secular stellar mass
loss is also important because it may in certain contexts be an
important source of star formation fuel during late stages of galaxy
evolution \citep[e.g.,][]{Leitner11, Voit11}, and may also feed the
central black hole \citep[e.g.,][]{Ciotti01}.  Standard implementations of
stellar mass loss in galaxy formation simulations include only an
instantaneous return of mass to the surrounding gas after a star
formation event \citep[e.g.,][]{Katz96}.  This does not capture mass
loss and heating due to old stellar populations discussed
here. Although secular stellar mass loss is now increasingly included
in galaxy formation simulations \citep[e.g.,][]{Kravtsov05, Stinson06,
  Agertz13, Vogelsberger13}, current simulations only account for the
mass and momentum transfer of the ejecta, but not the heating due to
dissipation of its kinetic energy.  Although implementation of such
heating in numerical simulations is, in principle, straightforward,
care must be taken to ensure that energy injected due to the mass loss
heating is not dissipated away excessively due to resolution effects.

The AGB heating scenario has a variety of observational implications.
Stripping of the circumstellar shells should produce anti-correlations
between the specific frequency of planetary nebulae and galaxy
velocity dispersion, both within and among galaxies. This is supported
by currently existing data \citep{Coccato09} and is a strong
prediction of the model, although the timescale over which the
stripping takes place is uncertain and will determine whether or not
the visibility of planetary nebulae is affected.  As the mass loss
rate decreases with stellar age, we expect the oldest galaxies to have
the most difficulty maintaining low levels of star formation, and
perhaps some galaxies will show evidence of recent rejuvenation.  As
discussed in the previous section, AGB heating in S0 galaxies will be
less effective than in true ellipticals.  In addition, we expect SFR
activity to be correlated with a {\it combination} of stellar density
and velocity dispersion and so outliers in the $\rho_\ast-\sigma_\ast$
plane should provide valuable insight.  Nearby environments such as the
bulge of M31 and Centaurus A may provide important constraints on the
interaction between ejected stellar envelopes and the surrounding hot
diffuse gas.  Heating caused by the thermalization of stellar ejecta
may be responsible for the close similarity between the specific
kinetic energy of the stars and the specific internal energy of the
ambient gas \citep[e.g.,][]{Nagino09}.

%---------------------------------------------------------%

\acknowledgments 

We thank the referee for very helpful comments that improved the
quality and clarity of the manuscript.  We also acknowledge useful
discussions with Luca Ciotti and Jerry Ostriker.  CC is supported by
Packard and Sloan Foundation Fellowships.  CC and PvD acknowledge the
Lorentz Center meeting "What Regulates Galaxy Evolution?" in April
2013 where this work was conceived.

%\bibliography{../master_refs}

%\input{ms.bbl}

\end{document}